\documentclass[journal]{llncs}

\usepackage{amsfonts}
\usepackage{amssymb}
\usepackage{stfloats}
\usepackage{cite}
\usepackage{caption}
\usepackage{graphicx}
\usepackage{psfrag}
\usepackage{subfigure}
\usepackage{amsmath}
\usepackage{array}
\usepackage{algorithm}
\usepackage{algpseudocode}
\usepackage[affil-it]{authblk}
\usepackage[english]{babel}
\usepackage{blindtext}
\usepackage{url}
\usepackage{epstopdf}
\usepackage{flushend}
\usepackage{multirow}
\usepackage{bm}

\title{Attention based Convolutional Recurrent Neural Network for Environmental Sound Classification}

\author{Zhichao Zhang\and Shugong Xu\thanks{Corresponding author.
Shanghai Institute for Advanced Communication and Data Science,
Shanghai University, Shanghai, China(email: \email{shugong@shu.edu.cn}).}\and Tianhao Qiao\and Shunqing Zhang\and Shan Cao}
\institute{Shanghai Institute for Advanced Communication and Data Science,\\
Shanghai University, Shanghai, 200444, China\\
\email{\{zhichaozhang, shugong, qiaotianhao, shunqing, cshan\}@shu.edu.cn}}

\begin{document}

\maketitle

\begin{abstract}
Environmental sound classification (ESC) is a challenging problem due to the complexity of sounds. The ESC performance is heavily dependent on the effectiveness of representative features extracted from the environmental sounds. However, ESC often suffers from the semantically irrelevant frames and silent frames. In order to deal with this, we employ a frame-level attention model to focus on the semantically relevant frames and salient frames. Specifically, we first propose an convolutional recurrent neural network to learn spectro-temporal features and temporal correlations. Then, we extend our convolutional RNN model with a frame-level attention mechanism to learn discriminative feature representations for ESC. Experiments were conducted on ESC-50 and ESC-10 datasets. Experimental results demonstrated the effectiveness of the proposed method and achieved the state-of-the-art performance in terms of classification accuracy.


\end{abstract}


\section{Introduction}

Environmental sound classification (ESC) is an important branch of sound recognition and is widely applied in surveillance\cite{radhakrishnan2005audio}, home automation\cite{vacher2007sound}, scene analysis\cite{barchiesi2015acoustic} and machine hearing\cite{lyon2010machine}.

Thus far, a variety of signal processing and machine learning techniques have been applied for ESC, including dictionary learning\cite{chu2009environmental}, matrix factorization\cite{bisot2017feature}, gaussian mixture model (GMM)\cite{dhanalakshmi2011classification} and recently, deep neural networks\cite{zhang2018deep,salamon2017deep}. For traditional machine learning classifiers, selecting proper features is key to effective performance. For instance, audio signals have been traditionally characterized by Mel-frequency cepstral
coefficients (MFCCs) as features and classified using a GMM classifier.

In recent years, deep neural networks (DNNs) have shown outstanding performance in feature extraction for ESC. Compared to hand-crafted feature, DNNs have the ability to extract discriminative feature representations from large quantities of
training data and generalize well on unseen data. McLoughlin et al.\cite{mcloughlin2015robust} proposed a deep belief network
to extract high-level feature representations from magnitude spectrum which yielded better results than the traditional methods. Piczak\cite{piczak2015environmental} first evaluated the potential of convolutional neural network (CNN) in classifying short audio clips of environmental sounds and showed excellent performance on several public datasets. Takahashi et al.\cite{takahashi2016deep} created a three-channel feature as the input to a CNN by combining log mel spectrogram and its delta and delta-delta information in a manner similar to the RGB input of image. In order to model the sequential dynamics of environmental sound signals, Vu et al.\cite{vu2016acoustic} applied a recurrent neural network (RNN) to learn temporal relationships. Moreover, there is a growing trend to combine CNN and RNN models into a single architecture. Bae et al.\cite{bae2016acoustic} proposed to train the RNN and CNN in parallel in order to learn sequential correlation and local spectro-temporal information.


In addition, attention mechanism-based models have shown outstanding performance in learning relevant feature representations for sequence data\cite{chorowski2015attention}. Recently, attention mechanism-based RNNs have been successfully applied to a wide variety of tasks, including speech recognition\cite{chorowski2015attention}, machine translation\cite{bahdanau2014neural} and document classification\cite{yang2016hierarchical}. In principle, attention mechanism-based RNNs are well suited to ESC tasks. First, environmental sound is essentially the sequence data which contains correlation information between adjacent frames. Second, not all frame-level features contribute equally to the representations of environmental sounds. Usually, in public ESC datasets, signals contains many periods of silence, with only a few intermittent frames associated with the sound class. Thus, it is important to select semantically relevant frames for specific class. Similar to attention mechanism-based RNN, we can also compute the frame-level attention map from CNN features, focusing on the semantically relevant frames. In the field of ESC, several works\cite{WJ2018ASC,ZR2018ASC,li2019multi,guo2017attention} have studied the effectiveness of attention mechanisms and have obtained promising results in several datasets. Different from previous works, we explored both the performance of frame-level attention mechanism for CNN layers and RNN layers.

In this paper, we propose an attention mechanism-based convolutional RNN architecture (ACRNN) in order to focus on semantically relevant frames and produce discriminative features for ESC. The main contributions of this paper are summarized as follows.

\begin{itemize}
    \item  To deal with silent frames and semantically irrelevant frames, We employ an attention model to automatically focus on the semantically relevant frames and produce discriminative features for ESC. We explore both the performance of frame-level attention mechanism for CNN layers and RNN layers.
    \item  To analyze temporal relations, We propose a novel convolutional RNN model which first uses CNN to extract high level feature representations and then inputs the features to bidirectional GRUs. We combine the convolutional RNN and attention model in a unified architecture.
    \item  To indicate the effectiveness of the proposed method and achieve current state-of-the-art performance, we conduct experiments on ESC-10 and ESC-50 datasets.
\end{itemize}


\begin{figure}[ht]
\centering
        \includegraphics[width=4.8 in]{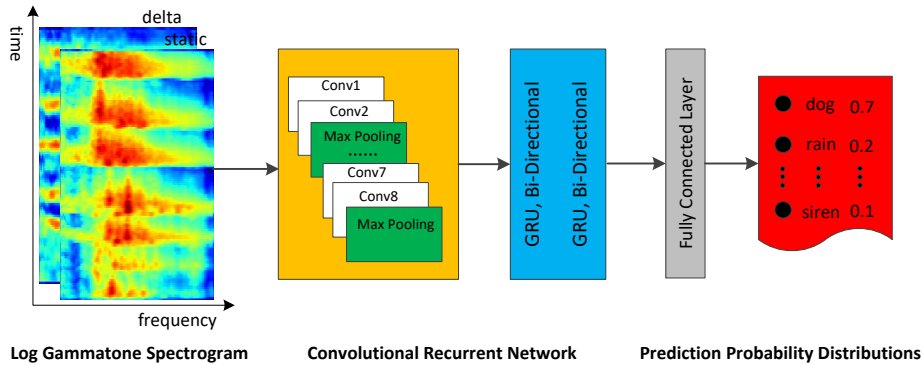}
        \caption{Architecture of convolutional recurrent neural network for environmental sound classification}
        \label{fig:framework}
\end{figure}

\section{Methods} \label{sect:methods}
In this section, we introduce the proposed method for ESC. First, we generate Log Gammatone spetrogram (Log-GTs) features from environmental sounds as the input of ACRNN, as shown in Fig. \ref{fig:framework}. Then, we introduce the architecture of ACRNN, which combines convolutional RNN and a frame-level attention mechanism. For the architecture of convolutional RNN and attention mechanism, we will give a detailed description, respectively. Finally, the data augmentation methods we used are introduced.

\subsection{Feature Extraction and Preprocessing}
Given a signal, We first use short-time Fourier Transform (STFT) with hamming window size of 23 ms (1024 samples at 44.1kHz) and 50$\%$ overlap to extract the energy spectrogram. Then, we apply a 128-band Gammatone filter bank\cite{valero2012gammatone} to the energy spectrogram and the resulting spectrogram is converted into logarithmic scale. In order to make efficient use of limited data, the spectrogram is split into 128 frames (approximately 1.5s in length) with 50$\%$ overlap. The delta information of the original spectrogram is calculated, which is the first temporal derivative of the static spectrogram. Afterwards, we concatenate the log gammatone spectrogram and its delta information to a 3-D feature representation $X\in{R^{128\times{128}\times{2}}}$ (Log-GTs) as the input of the network.

\subsection{Architecture of Convolutional RNN}
In this section, we propose an convolutional RNN to analyze Log-GTs for ESC. We first use CNN to learn high level feature representations on the Log-GTs. Then, the CNN-learned features are fed into bidirectional gated recurrent unit (GRU) layers which are used to learn the temporal correlation information. Finally, these features are fed into a fully connected layer with a softmax activation function to output the probability distribution of different classes.
In this paper, the convolutional RNN is comprised of eight convolutional layers (\emph{$l_1$}-\emph{$l_8$}) and two bidirectional GRU layers (\emph{$l_9$}-\emph{$l_{10}$}). The architecture and parameters of network are as follows:
\begin{itemize}
    \item  \emph{$l_1$}-\emph{$l_2$}: The first two stacked convolutional layers use 32 filters with a receptive field of (3,5) and stride of (1,1). This is followed by a max-pooling with a (4,3) stride to reduce the dimensions of feature maps. ReLU activation function is used.
    \item  \emph{$l_3$}-\emph{$l_4$}: The next two convolutional layers use 64 filters with a receptive field of (3,1) and stride of (1,1), and is used to learn local patterns along the frequency dimension. This is followed by a max-pooling with a (4,1) stride. ReLU activation function is used.
    \item  \emph{$l_5$}-\emph{$l_6$}: The following pair of convolutional layers uses 128 filters with a receptive field of (1,5) and stride of (1,1), and is used to learn local patterns along the time dimension. This is followed by a max-pooling with a (1,3) stride. ReLU activation function is used.
    \item  \emph{$l_7$}-\emph{$l_8$}: The subsequent two convolutional layers use 256 filters with a receptive field of (3,3) and stride of (1,1) to learn joint time-frequency characteristics. This is followed by a max-pooling of a (2,2) stride. ReLU activation function is used.
    \item  \emph{$l_9$}-\emph{$l_{10}$}: Two bidirectional GRU layers with 256 cells are used for temporal summarization, and tanh activation function is used. Dropout with probability of $0.5$ is used for each GRU layer to avoid overfitting.
\end{itemize}

Batch normalization\cite{ioffe2015batch} is applied to the output of the convolutional layers to speed up training. L2-regularization is applied to the weights of each layer with a coefficient $0.0001$.


\subsection{Frame-level Attention Mechanism}

\begin{figure}[ht]
\centering
        \includegraphics[width=4.8 in]{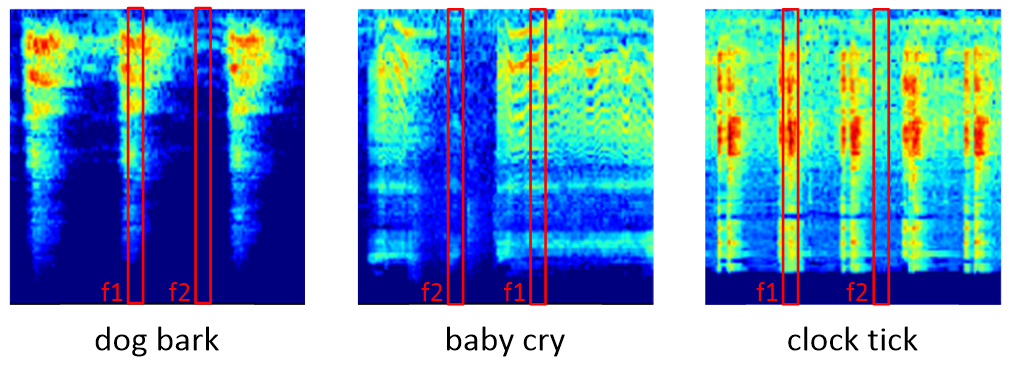}
        \caption{Visualization of Log-GTs of different classes with semantically relevant frame($f1$) and silent or noisy frame($f2$). From left to right, the class is \emph{dog bark}, \emph{baby cry} and \emph{clock tick}. }
        \label{fig:framelevel_att}
\end{figure}

Not all frame-level features contribute equally to representations of environmental sounds. As shown in Fig. \ref{fig:framelevel_att}, except for the semantically relevant frames($f1$), the features usually contain silent or noisy frames($f2$), which reduce the robustness of model and increase misclassification. Hence, we apply frame-level attention mechanisms to focus on the parts that are most vital to the meaning of the sound and to produce discriminative representations for ESC. In this paper, we employ attention mechanism for CNN layers and RNN layers, respectively.

\subsubsection{Attention for CNN layers:}

As shown in Fig.\ref{fig:cnn_rnn_attention}(a), given CNN features $M\in{R^{F\times{T}\times{C}}}$, we first use a 3x3 convolution filter to learn a hidden representation. 
This is followed by a average-pool with $(F,1)$ size in order to reduce the frequency dimension to one. Then, we use softmax function to form a normalized attention map $A\in{R^{1\times{T}\times{1}}}$, which holds the frame-level attention weights for CNN features. With attention map $A$, the attention weighted CNN features are obtained as 

\begin{equation}\label{eqn:cnn_att}
M'=M\cdot{A}
\end{equation}

The attention is applied by multiplying the attention vector $A$ to each feature vector of $M$ along frequency dimension and channel dimension.

\subsubsection{Attention for RNN layers:}

\begin{figure}[t]
\centering
        \includegraphics[width=4.7 in]{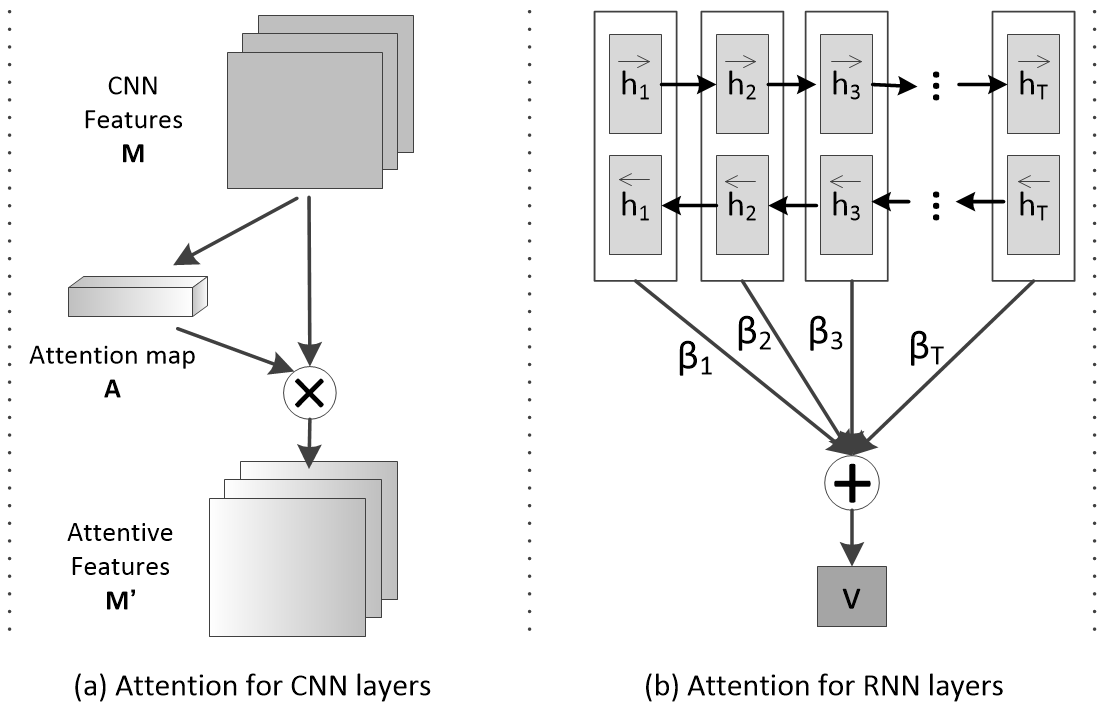}
        \caption{Frame-level attention for (a) CNN layers and (b) RNN layers. For CNN layers, we use frame-level attention to obtain attention map, which is multiplied in frame-wise of CNN features, resulting the attention weighted features. For RNN layers, we utilize frame-level attention to obtain attention weights, which is multiplied in frame-wise of input features. Then, we aggregate these attention weighted representations to form a feature vector, which can be seen as a high-level representation of a sound like "dog bark".}
        \label{fig:cnn_rnn_attention}
\end{figure}

As shown in Fig.\ref{fig:cnn_rnn_attention}(b), we first feed the GRU output $h_t=[\overrightarrow{h_t}, \overleftarrow{h_t}]$ through a one-layer MLP to obtain a hidden representation of $h_t$, then we calculate the normalized importance weight $\beta_t$ by a softmax function (\ref{eqn:norm}). After that, we compute the feature vector $v$ through a weighted sum of the frame-level convolutional RNN feautues based on the weights (\ref{eqn:sum}). The feature vector $v$ is forwarded into the fully connected layer for final classification.


\begin{equation}\label{eqn:norm}
\beta_t=\frac{exp(W*h_t)}{\sum_{t=1}^T{exp(W*h_t)}}
\end{equation}

\begin{equation}\label{eqn:sum}
v=\sum_{t=1}^T{\beta_t{h_t}}
\end{equation}

\subsection{Data Augmentation}
\label{subsect:mixup}
Limited data easily leads model towards overfitting. In this paper, we use time stretch with a factor randomly selected from [0.8, 1.3] and pitch shift with a factor randomly selected from [-3.5, 3.5] to increase raw training data size. In addition, an efficient mixup \cite{zhang2017mixup} augmentation method is used to construct virtual training data and extend the training distribution. In mixup, a feature and a target (\^x, \^y) are generated by mixing two feature-target examples, which are determined by

\begin{equation}\label{eqn:mapping}
\left\{
\begin{aligned}
\hat{\mathbf x}= {\lambda}x_i + (1-\lambda)x_j\\
\hat{\mathbf y}= {\lambda}y_i + (1-\lambda)y_j
\end{aligned}
\right.
\end{equation}

where $x_i$ and $x_j$ are two features randomly selected from the training Log-GTs, and $y_i$ and $y_j$ are their one-hot labels. The mix factor $\lambda$ is decided by a hyper-parameter $\alpha$ and $\lambda$ $\sim$ Beta($\alpha$, $\alpha$).

\section{Experiments and Results} \label{sect:exp}

\subsection{Experiment Setup}

To evaluate the performance of our proposed methods, we carry out experiments on two publicly available datasets: ESC-50 and ESC-10\cite{piczak2015esc}. ESC-50 is a collection of 2000 environmental recordings containing 50 classes in 5 major categories, including \emph{animals, natural soundscapes and water sounds, human non-speech sounds, interior/domestic sounds}, and \emph{exterior/urban noises}. All audio samples are 5 seconds in duration with a 44.1 kHz sampling frequency. ESC-10 is a subset of 10 classes (400 samples) selected from the ESC-50 dataset (\emph{dog bark, rain, sea waves, baby cry, clock tick, person sneeze, helicopter, chainsaw, rooster, fire crackling}).

In this paper, we use a sampling rate of 44.1 kHz for all samples in order to use rich high-frequency information. For training, all models optimize cross-entropy loss using mini-batch stochastic gradient descent with Nesterov momentum of 0.9. Each batch consists of 64 segments randomly selected from the training set without repetition. All models are trained for 300 epochs by beginning with an initial learning rate of 0.01, and then divided the learning rate by 10 every 100 epochs. We initialize the network weights to zero mean Gaussian noise with a standard deviation of 0.05. In the test phase, we evaluate the whole sample prediction with the highest average prediction probability of each segment. Both the training and testing features are normalized by the global mean and stardard deviation of the training set. All models are trained using Keras library with TensorFlow backend on a Nvidia P100 GPU with 12GB memory.

\subsection{Experiment Results}

\begin{table}[ht]
\caption{Comparison of ACRNN and existing methods. We perform 5-fold cross validation (CV) by using the official fold settings. The average results of CV are recorded.
\label{tab:res}}
\centering  %
\setlength{\tabcolsep}{7mm}{
\begin{tabular}{lcccccccc}
\hline
\hline
\textbf{Model} &\textbf{ESC-10} &\textbf{ESC-50}\\
\hline
PiczakCNN\cite{piczak2015environmental} &80.5\% &64.9\%\\
\hline
SoundNet\cite{aytar2016soundnet} &92.1\% &74.2\%\\
\hline
WaveMsNet\cite{zhu2018learning} &93.7\% &79.1\%\\
\hline
EnvNet-v2\cite{tokozume2017learning} &91.4\% &84.9\%\\
\hline
Multi-Stream CNN\cite{li2019multi} &93.7\% &83.5\%\\
\hline
ACRNN &\bf{93.7\%} &\bf{86.1\%}\\
\hline
\hline
\end{tabular}}
\end{table}

We compare our model with existing networks reported as PiczakCNN\cite{piczak2015environmental}, SoundNet\cite{aytar2016soundnet}, WaveMsNet\cite{zhu2018learning}, EnvNet-v2\cite{tokozume2017learning} and Multi-Stream CNN\cite{li2019multi}. According to \cite{piczak2015environmental}, PiczakCNN consists of two convolutional layers and three fully connected layers. The input features of CNN are generated by combining log mel spectrogram and its delta information. We refer PiczakCNN as a baseline method. 






The results are summarized in Table \ref{tab:res}. We see that ACRNN outperforms PiczakCNN and obtains an absolute improvement of 13.2$\%$ and 21.2$\%$ on ESC-10 and ESC-50 datasets, respectively. Then, we compare our model with several state-of-the-art methods: SoundNet8\cite{aytar2016soundnet}, WaveMsNet\cite{zhu2018learning}, EnvNet-v2\cite{tokozume2017learning} and Multi-Stream CNN\cite{li2019multi}. We observe that on both ESC-10 and ESC-50 datasets, ACRNN obtains the highest classification accuracy. Note that WaveMsNet\cite{zhu2018learning} and Multi-Stream CNN\cite{li2019multi} achieve same classification accuracy as ACRNN on ESC-10 but using feature fusion (raw data and spectrogram features), whereas ACRNN only utilizes spectrogram features.

\begin{figure}[ht]
\centering
        \includegraphics[width=4.6 in]{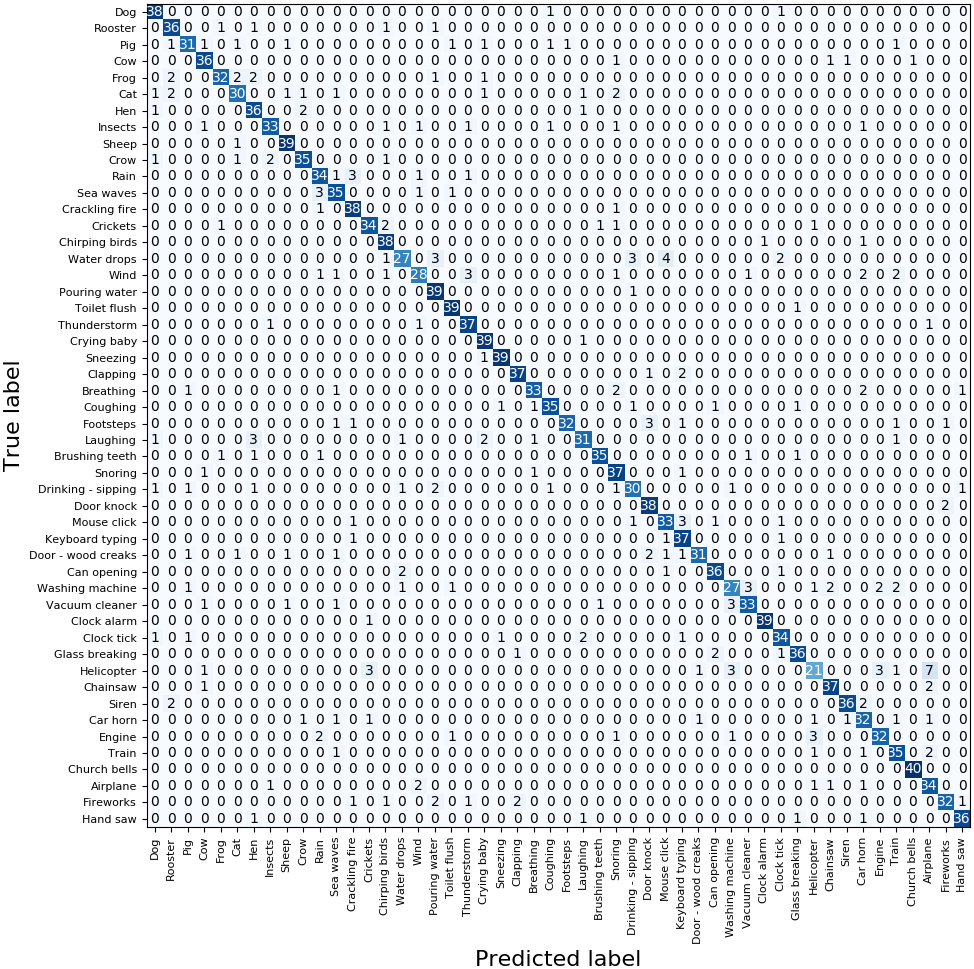}
        \caption{Confusion matrix of ACRNN with an average classification accuracy 86.1$\%$ on ESC-50 dataset.}
        \label{fig:conf_esc50}
\end{figure}

In Fig.\ref{fig:conf_esc50}, we provide the confusion matrix generated by ACRNN for ESC-50 dataset. We see that most classes achieve higher accuracy than 80$\%$(32/40). Particularly, \emph{Church bells} obtains a 100$\%$ recognition rate. However, we observe that only 52.5$\%$(21/40) \emph{Helicopter} samples are correctly recognized, with 17.5$\%$(7/40) samples misclassified as \emph{Airplane}. We attribute this mistakes to the similar characteristics between the two environmental sounds.

\subsection{Effects of attention mechanism}

\begin{table}[h]
\caption{Classification accuracy of proposed convolutional RNN with and without the attention mechanism. 'augment' denotes a combination of time stretch, pitch shift and mixup.
\label{tab:attention}}
\centering  %
\setlength{\tabcolsep}{4mm}{
\begin{tabular}{lcccccccc}
\hline
\hline
\textbf{Model Settings} &\textbf{ESC-10} &\textbf{ESC-50}\\
\hline
convolutional RNN &89.2\% &79.9\%\\
\hline
convolutional RNN-attention &91.7\% &81.3\%\\
\hline
convolutional RNN-augment &93.0\% &84.6\%\\
\hline
convolutional RNN-attention-augment &\textbf{93.7\%} &\textbf{86.1\%}\\
\hline
\hline
\end{tabular}}
\end{table}

To investigate the effects of the attention mechanism, we compare the results of proposed convolutional RNN with and without the attention mechanism. In Table \ref{tab:attention}, the results show that the attention mechanism delivers a significantly improved accuracy even when we use a data augmentation scheme. In addition, data augmentation boasts an improvement of 2.0$\%$ and 4.8$\%$ on ESC-10 and ESC-50 datasets, respectively.

\subsection{Where to apply attention}

\begin{table}[h]
\caption{Classification accuracy of applying the attention mechanism to the output of different layers of the proposed convolutional RNN.
\label{tab:cnn_attention}}
\centering  %
\setlength{\tabcolsep}{8mm}{
\begin{tabular}{lcccccccc}
\hline
\hline
\textbf{Model Settings} &\textbf{ESC-10} &\textbf{ESC-50}\\
\hline
no attention &93.0\% &84.6\%\\
\hline
attention at \emph{$l_2$} &93.5\% &85.2\%\\
\hline
attention at \emph{$l_4$} &92.7\% &83.8\%\\
\hline
attention at \emph{$l_6$} &92.7\% &84.4\%\\
\hline
attention at \emph{$l_8$} &92.5\% &84.9\%\\
\hline
\hline
attention at \emph{$l_{10}$} &\textbf{93.7\%} &\textbf{86.1\%}\\
\hline
\hline
\end{tabular}}
\end{table}

In this section, we investigate the classification performance when applying frame-level attention mechanism to the different layers of CNN and RNN. As shown in Table \ref{tab:cnn_attention}, we obtained the highest classification accuracy and boosted an absolutely improvement of 0.7\% and 1.5\% when applying the attention mechanism at \emph{$l_{10}$} on both ESC-10 and ESC-50 datasets, respectively. On the ESC-50 dataset, the classification accuracy obtained a slight improvement when the attention mechanism was applied at $l_2$ and $l_8$, while for other CNN layers, the classification accuracy decreased. On the ESC-10 dataset, we obtained an improvement of 0.5\% when only applying attention at $l_2$ for CNN layers. Furthermore, we found that on both ESC-10 and ESC-50 datasets, the classification accuracy is improved than standard convolutional RNN when applying attention at $l_2$ for CNN layers. 

\section{Conclusion} \label{sect:conc}
In this paper, we proposed an attention mechanism-based convolutional recurrent neural network (ACRNN) for ESC. We explored the frame-level attention mechanism and gave a detailed description for CNN layers and RNN layers, respectively. Experimental results on ESC-10 and ESC-50 datasets demonstrated the effectiveness of the proposed method and achieved state-of-the-art performance in terms of classification accuracy. In addition, we compared the classification accuracy when applying different layers, including CNN layers and RNN layers. The experimental results showed that applying attention for RNN layers obtained highest accuracy. However, we found when applying attention for CNN layers, the performance is not always improved. We plan to explore this in our future work.

\bibliographystyle{splncs04}
\bibliography{prcv19_esc}
\end{document}